# Full vectoring optimal power allocation in xDSL channels under per-modem power constraints and spectral mask constraints

Vincent Le Nir, Marc Moonen, Jan Verlinden, Mamoun Guenach

*Abstract*— In xDSL systems, crosstalk can be separated into two categories, namely in-domain crosstalk and out-of-domain crosstalk. In-domain crosstalk is also refered to as self crosstalk. Out-of-domain crosstalk is crosstalk originating from outside the multi-pair system and is also denoted as external noise (alien crosstalk, radio frequency interference,...). While self crosstalk in itself can easily be canceled by a linear detector like the ZF detector, the presence of external noise requires a more advanced processing. Coordination between transmitters and receivers enables the self crosstalk and the external noise to be mitigated using MIMO signal processing, usually by means of a whitening filter and SVD. In this paper, we investigate the problem of finding the optimal power allocation in MIMO xDSL systems in the presence of self crosstalk and external noise. Optimal Tx/Rx structures and power allocation algorithms will be devised under practical limitations from xDSL systems, namely per-modem total power constraints and/or spectral mask constraints, leading to a generalized SVD-based transmission. Simulation results are given for bonded VDSL2 systems with external noise coming from ADSL2+ or VDSL2 disturbing lines, along with a comparison between algorithms with one-sided signal coordination either only at the transmit side or the receive side.

## I. INTRODUCTION

The growing demand for high speed services in access networks calls for new paradigms offering an increased capacity and better performance. Thanks to the success of x-Digital Subscriber Lines (xDSL) and ADSL in particular, service providers begin to bind copper pairs, allowing customers to be served with higher bitrates through the usage of adequate Multiple Input Multiple Output (MIMO) signal processing algorithms. This processing is also to provide a suitable crosstalk interference mitigation. In xDSL systems, crosstalk can be separated into two categories, namely in-domain crosstalk and out-of-domain crosstalk. In-domain crosstalk is also refered to as self crosstalk. Out-of-domain crosstalk is crosstalk originating from outside the multi-pair system and is also denoted as external noise (alien crosstalk, radio frequency interference,...).

Self crosstalk cancellation has been studied for two-sided coordination vector channels and for one-sided coordination Multiple Access Channels (MAC) or Broadcast Channels (BC). For two-sided coordination vector channels, the Channel State Information (CSI) is available at both the transmitter and the receiver. For this full vectoring problem, the optimal precoding at the transmitter and equalization at the receiver as well as optimal Power Spectral Densities (PSD's) are obtained through the SVD of the channel matrix combined with standard waterfilling [1]. For one-sided coordination MAC or BC, only the receivers for the MAC or the transmitters for the BC can cooperate. It has been shown that the optimal structure for MAC is a Minimum Mean Square Error-Decision Feedback Equalizer (MMSE-DFE) along with a power allocation found by exhaustive search [2]. For BC, a similar (dual) optimal structure has also been desribed called MMSE-Dirty Paper Coding (MMSE-DPC). Moreover, owing to the diagonal dominance structure of the channel matrix, all these optimal structures can be simplified to Zero Forcing (ZF) solution for the MAC (or a simple Diagonalizing Precoder (DP) for the BC) with transmit PSD's obtained by single-user waterfilling [3], [4]. Finally, when there is no coordination, neither at the receive side nor at the transmit side, this leads to Interference Channels (IC) where spectral management is employed to reduce crosstalk. For IC, the optimal transmit PSD's have been found by means of Optimal Spectrum Balancing (OSB) [5], [6].

External noise is most often the predominant interferer and originates from outside the multi-pair system. With external noise, the diagonal dominance structure of the channel matrix is destroyed by the necessary whitening, and hence the simple ZF solution along with single-user waterfilling is found to be suboptimal. The basic idea of external noise cancellation is then to exploit the correlation of the noise to improve the performance of the transmission and hence to increase the total capacity. This noise correlation can appear in the spatial domain (between pairs), the frequency domain (between tones) or the mode domain (between common-mode and differential-mode) [7]. In a recent paper [8], it was shown that there is more benefit in exploiting the noise correlation between pairs rather than the correlation between tones.

In this paper, we investigate the problem of finding the optimal power allocation in MIMO xDSL systems under self crosstalk and external noise and with two-sided coordination exploiting the noise correlation between pairs. Coordination between transmitters and receivers enables the self crosstalk and the external noise to be mitigated using MIMO signal processing by means of a whitening filter and Singular Value Decomposition (SVD). Optimal transmitter/receiver (Tx/Rx) structures and power allocation algorithms will be devised with practical limitations from xDSL systems, namely per-modem total power constraints and spectral mask constraints, leading to a generalized SVD-based transmission. Compared to [8] where the external noise is mitigated under a total power constraint without self crosstalk, the proposed algorithms mitigate the external noise under per-modem total power constraints and spectral mask constraints considering the self crosstalk. Capitalizing on the results of [8] where it is shown that there is more correlation in the spatial domain than the frequency domain, we consider the correlation only in the spatial domain. Contrary to [7] where CSI is available only at the receive side and the external noise is mitigated by common-mode exploitation, our algorithms assume CSI at the transmit and the receive sides.

In section II, we first recall the optimal power allocation for two-sided coordination vector channels (i.e. full vectoring) with self crosstalk and external noise under a total power constraint. The primal MIMO capacity optimization problem subject to a total power constraint coupled over the tones is transformed into a collection of per-tone unconstrained optimization problems using a dual problem formulation. We derive optimal transmitter and receiver structures (precoders and equalizers) in combination with power allocation which achieve MIMO channel capacity. Secondly, we devise the optimal power allocation for two-sided coordination vector channels with self crosstalk and external noise under per-modem total power constraints and spectral mask constraints, leading to a generalized SVD-based transmission (section III). Similar derivations are given for the special (or simpler) per-modem total power constraints case. Simulation results are given for bonded VDSL2 systems with external noise coming from ADSL2+ or VDSL2 disturbing lines, along with a comparison between algorithms with one-sided signal coordination either only at the transmit side or the receive side (section IV).

V. Le Nir and M. Moonen are with the SISTA/ESAT laboratory, Katholieke Universiteit Leuven, Leuven, Belgium. E-mail: vincent.lenir@esat.kuleuven.be marc.moonen@esat.kuleuven.be

Jan Verlinden and M. Guenach are respectively with Alcatel-Lucent and Alcatel-Lucent Bell Labs Antwerpen.

## II. TOTAL POWER CONSTRAINT

In this paragraph, we recall the classical SVD-based algorithm with optimal power allocation for two-sided coordination vector channels with self crosstalk and external noise under a total power constraint. We assume that transmitters use Discrete Multi-Tone (DMT) modulation with a cyclic prefix longer than the maximum delay spread of the channel. As shown in [8], there is more correlation in the spatial domain than the frequency domain. In this paper, we exploit the noise correlation in the spatial domain and we assume that the external noise is synchronized with the MIMO system. Therefore, there is no correlation in the frequency domain and the external noise is decoupled over the tones. The transmission over one tone can then be modelled as:

$$\mathbf{y}_i = \mathbf{H}_i \mathbf{x}_i + \mathbf{n}_i \quad i = 1 \ldots N_c \tag{1}$$

where $N_c$ is the number of subcarriers, $\mathbf{x}_i$ is the vector of $N$ transmitted signals on tone $i$, $\mathbf{y}_i$ the received signal vector, $\mathbf{H}_i$ the $N \times N$ MIMO channel matrix and $\mathbf{n}_i$ the vector of noise containing Additive White Gaussian Noise (AWGN) and external noise (alien crosstalk, radio frequency interference,...). The primal problem of finding optimal PSD's for a MIMO binder under a total power constraint $P^{tot}$ is:

$$\begin{aligned} \max_{(\boldsymbol{\Phi}_i)_{i=1\ldots N_c}} & \quad C(\boldsymbol{\Phi}_i)_{i=1\ldots N_c} \\ \text{subject to} & \sum_{i=1}^{N_c} Trace(\boldsymbol{\Phi}_i) \leq P^{tot} \\ & \boldsymbol{\Phi}_i \succeq 0, i = 1 \ldots N_c \end{aligned} \tag{2}$$

with $\boldsymbol{\Phi}_i$ the covariance matrix of transmitted symbols $\boldsymbol{\Phi}_i = E[\mathbf{x}_i \mathbf{x}_i^H]$ over tone i for the MIMO binder and with the objective function being the MIMO capacity summed over the $N_c$ tones [9]:

$$C(\boldsymbol{\Phi}_i)_{i=1\ldots N_c} = \sum_{i=1}^{N_c} log_2 \left[ det \left( \mathbf{I} + \mathbf{H}_i \boldsymbol{\Phi}_i \mathbf{H}_i^H \mathbf{R}_i^{-1} \right) \right] \tag{3}$$

Here, $\mathbf{R}_i$ is the covariance matrix of the noise $\mathbf{R}_i = E[\mathbf{n}_i \mathbf{n}_i^H]$. The idea of dual decomposition is to solve (2) via its Lagrangian [10]. The Lagrangian decouples into a set of $N_c$ smaller problem, thus reducing the complexity of equation (2). The dual objective function is:

$$F(\lambda) = \max_{(\boldsymbol{\Phi}_i)_{i=1\ldots N_c}} \mathcal{L}(\lambda, (\boldsymbol{\Phi}_i)_{i=1\ldots N_c}) \tag{4}$$

with

$$\mathcal{L}(\lambda, (\boldsymbol{\Phi}_i)_{i=1\ldots N_c}) = \sum_{i=1}^{N_c} \left( log_2 \left[ det \left( \mathbf{I} + \mathbf{H}_i \boldsymbol{\Phi}_i \mathbf{H}_i^H \mathbf{R}_i^{-1} \right) \right] - \lambda Trace(\boldsymbol{\Phi}_i) \right) + \lambda P^{tot} \tag{5}$$

with $\lambda$ the Lagrange multiplier. The dual optimization problem is:

$$\begin{aligned} \min_{\lambda} & \quad F(\lambda) \\ \text{subject to} & \quad \lambda \geq 0 \end{aligned} \tag{6}$$

Because the dual function is convex in $\lambda$, standard convex optimization results guarantee that the primal problem (2) and the dual problem (6) have the same solution [11]. Indeed, the objective and constraint functions are differentiable and the Slater's conditions are satisfied, therefore the duality gap is zero and the minimum of the dual function corresponds to the global optimum of the primal problem [10]. The search for the optimal $\lambda$ in (6) involves evaluations of the dual objective function (4), i.e. maximizations of the Lagrangian, which is decoupled over the tones for the given $\lambda$. By exploiting the Cholesky decomposition $\mathbf{R}_i = \mathbf{L}_i \mathbf{L}_i^H$, where $\mathbf{L}_i$ is a lower triangular matrix (whose inverse will be used to whiten the noise at the receive side), we obtain the following equation (using the property $det(\mathbf{I} + \mathbf{AB}) = det(\mathbf{I} + \mathbf{BA})$):

$$\mathcal{L}(\lambda, (\boldsymbol{\Phi}_i)_{i=1\ldots N_c}) = \sum_{i=1}^{N_c} \left( log_2 \left[ det \left( \mathbf{I} + \mathbf{L}_i^{-1} \mathbf{H}_i \boldsymbol{\Phi}_i \mathbf{H}_i^H \mathbf{L}_i^{-H} \right) \right] - \lambda Trace(\boldsymbol{\Phi}_i) \right) + \lambda P^{tot} \tag{7}$$

The Singular Value Decomposition (SVD) of the withened channel $\mathbf{L}_i^{-1} \mathbf{H}_i = \mathbf{U}_i \mathbf{D}_i \mathbf{V}_i^H$ transforms the initial channel into a product between two unitary matrices $\mathbf{U}_i$, $\mathbf{V}_i^H$ and a diagonal matrix of singular values $\mathbf{D}_i$.

$$\mathcal{L}(\lambda, (\boldsymbol{\Phi}_i)_{i=1\ldots N_c}) = \sum_{i=1}^{N_c} \left( log_2 \left[ det \left( \mathbf{I} + \mathbf{U}_i \mathbf{D}_i \mathbf{V}_i^H \boldsymbol{\Phi}_i \mathbf{V}_i \mathbf{D}_i \mathbf{U}_i^H \right) \right] - \lambda Trace(\boldsymbol{\Phi}_i) \right) + \lambda P^{tot} \tag{8}$$

By setting $\tilde{\boldsymbol{\Phi}}_i = \mathbf{V}_i^H \boldsymbol{\Phi}_i \mathbf{V}_i$ we can rewrite this as:

$$\mathcal{L}(\lambda, (\tilde{\boldsymbol{\Phi}}_i)_{i=1\ldots N_c}) = \sum_{i=1}^{N_c} \left( log_2 \left[ det \left( \mathbf{I} + \mathbf{D}_i^2 \tilde{\boldsymbol{\Phi}}_i \right) \right] - \lambda Trace(\tilde{\boldsymbol{\Phi}}_i) \right) + \lambda P^{tot} \tag{9}$$

The off-diagonal elements in $\tilde{\boldsymbol{\Phi}}_i$ merely reduce the determinant owing to diagonal matrices $\mathbf{D}_i^2$'s and the property $det(\mathbf{I} + \mathbf{A}) \leq Trace(\mathbf{I} + \mathbf{A})$. Hence the optimal $\tilde{\boldsymbol{\Phi}}_i$ is diagonal. In order to find the maximum, we compute the derivative of the function:

$$\frac{d\mathcal{L}(\lambda, (\tilde{\boldsymbol{\Phi}}_i)_{i=1\ldots N_c})}{d\tilde{\boldsymbol{\Phi}}_i} = \frac{1}{ln(2)} diag \left[ \left( \mathbf{D}_i^{-2} + \tilde{\boldsymbol{\Phi}}_i \right)^{-1} \right] - \lambda \mathbf{I} = \mathbf{0} \tag{10}$$

The optimal $\boldsymbol{\Phi}_i$ is given by:

$$\boldsymbol{\Phi}_i = \mathbf{V}_i \left[ \frac{\mathbf{I}}{ln(2)\lambda} - \mathbf{D}_i^{-2} \right]^+ \mathbf{V}_i^H \tag{11}$$

where the $[.]^+$ operation is inserted in order to obtain positive semi-definite $\boldsymbol{\Phi}_i$'s in formula (2). This is the well-known closed form waterfilling solution for MIMO systems[1]. The optimal power allocation consists of finding the optimal Lagrange multiplier which meets the total power constraint according to (11). The complete algorithm for the optimal power allocation under a total power constraint is given in the Annex A. The optimal Tx structure is given by the precoding matrix $\mathbf{V}_i$ while the optimal Rx structure is given by the equalizer matrix $\mathbf{U}_i^H$ leading to parallel SISO systems as defined by:

$$\mathbf{U}_i^H \mathbf{L}_i^{-1} \mathbf{y}_i = \mathbf{D}_i \tilde{\mathbf{x}}_i + \mathbf{U}_i^H \mathbf{L}_i^{-1} \mathbf{n}_i \tag{12}$$

---

[1] For practical implementations, we introduce the SNR gap $\Gamma$ referred as the code gap in [12] which is the SNR multiplier required to achieve the target probability of error at the desired data rate. Considering the same $\Gamma$ for the different virtual channels, the optimal $\boldsymbol{\Phi}_i$ is given by:

$$\boldsymbol{\Phi}_i = \mathbf{V}_i \left[ \frac{\mathbf{I}}{ln(2)\lambda} - \Gamma \mathbf{D}_i^{-2} \right]^+ \mathbf{V}_i^H$$

with the optimal power allocation under a total power constraint driven by (11).

## III. PER-MODEM TOTAL POWER CONSTRAINTS AND SPECTRAL MASK CONSTRAINTS

In the xDSL context, it is more relevant to consider a constraint on the power of each modem separately instead of a constraint on the power for all modems together. DSL standardization often defines spectral masks that each transmitter has to satisfy as well as the total power that each transmitter can transmit. In this section, we devise the optimal power allocation for two-sided coordination vector channels with self crosstalk and external noise under per-modem total power constraints (i.e. a single total power constraint for all tones per line) and spectral mask constraints as well as their corresponding optimal Tx/Rx structures.

### A. Optimal power allocation

The primal problem of finding optimal PSD's for a MIMO binder under per-modem total power constraints $P_j^{tot}$ and spectral mask constraints is:

$$\max_{(\mathbf{\Phi}_i)_{i=1...N_c}} C(\mathbf{\Phi}_i)_{i=1...N_c}$$
$$\text{subject to} \sum_{i=1}^{N_c}[\mathbf{\Phi}_i]_{jj} \leq P_j^{tot} \; \forall j \quad (13)$$
$$[\mathbf{\Phi}_i]_{jj} \leq \phi_i^{mask,j} \; \forall i \; \forall j$$
$$\mathbf{\Phi}_i \succeq 0, i = 1 \ldots N_c$$

with the objective function being the MIMO capacity summed over the $N_c$ tones given by (3). Again, we can apply the idea of dual decomposition by decoupling the primal problem into $N_c$ smaller problems [10] considering the per-modem total power constraints and the spectral amsk constraints. The dual objective function is:

$$F(\mathbf{\Lambda}, \tilde{\mathbf{\Lambda}}_1, \ldots, \tilde{\mathbf{\Lambda}}_{N_c}) = \max_{(\mathbf{\Phi}_i)_{i=1...N_c}} \mathcal{L}(\mathbf{\Lambda}, \tilde{\mathbf{\Lambda}}_1, \ldots, \tilde{\mathbf{\Lambda}}_{N_c}, (\mathbf{\Phi}_i)_{i=1...N_c}) \quad (14)$$

with

$$\mathcal{L}(\mathbf{\Lambda}, \tilde{\mathbf{\Lambda}}_1, \ldots, \tilde{\mathbf{\Lambda}}_{N_c}, (\mathbf{\Phi}_i)_{i=1...N_c}) =$$
$$\sum_{i=1}^{N_c} \left( log_2 \left[ det \left( \mathbf{I} + \mathbf{H}_i \mathbf{\Phi}_i \mathbf{H}_i^H \mathbf{R}_i^{-1} \right) \right] \right.$$
$$\left. - Trace((\mathbf{\Lambda} + \tilde{\mathbf{\Lambda}}_i)\mathbf{\Phi}_i) \right) + Trace\left(\mathbf{\Lambda} diag(P_j^{tot})\right) \quad (15)$$
$$+ \sum_{i=1}^{N_c} Trace\left(\tilde{\mathbf{\Lambda}}_i diag(\phi_i^{mask,j})\right)$$

The Lagrange multipliers corresponding to the per-modem total power constraints are contained in the diagonal matrix $\mathbf{\Lambda} = diag(\lambda_1, \ldots, \lambda_N)$, the Lagrange multipliers corresponding to the spectral mask constraints for tone i are contained in the diagonal matrix $\tilde{\mathbf{\Lambda}}_i = diag(\tilde{\lambda}_{i1}, \ldots, \tilde{\lambda}_{iN})$. The diagonal matrix $diag(\phi_i^{mask,j}) = diag(\phi_i^{mask,1}, \ldots, \phi_i^{mask,N})$ corresponds to the spectral mask for user j and tone i. The dual optimization problem is:

$$\begin{array}{c} \text{minimize} \\ \mathbf{\Lambda}, \tilde{\mathbf{\Lambda}}_1, \ldots, \tilde{\mathbf{\Lambda}}_{N_c} \end{array} F(\mathbf{\Lambda}, \tilde{\mathbf{\Lambda}}_1, \ldots, \tilde{\mathbf{\Lambda}}_{N_c})$$
$$\text{subject to} \quad [\mathbf{\Lambda}]_{jj}, [\tilde{\mathbf{\Lambda}}_1]_{jj}, \ldots, [\tilde{\mathbf{\Lambda}}_{N_c}]_{jj} \geq 0 \quad \forall j \quad (16)$$

The dual function is convex in $\mathbf{\Lambda}, \tilde{\mathbf{\Lambda}}_1, \ldots, \tilde{\mathbf{\Lambda}}_{N_c}$, therefore standard convex optimization results guarantee that the primal problem (13) and the dual problem (16) have the same solution [11]. The Lagrangian is differentiable and the Slater's conditions are satisfied, therefore the duality gap is zero and the minimum of the dual function corresponds to the global optimum of the primal problem [10]. The search for the optimal $\mathbf{\Lambda}, \tilde{\mathbf{\Lambda}}_1, \ldots, \tilde{\mathbf{\Lambda}}_{N_c}$ in (16) involves evaluations of the dual objective function (14), i.e. maximizations of the Lagrangian, which is decoupled over the tones for the given matrices $\mathbf{\Lambda}, \tilde{\mathbf{\Lambda}}_1, \ldots, \tilde{\mathbf{\Lambda}}_{N_c}$. By exploiting the Cholesky decomposition $\mathbf{R}_i = \mathbf{L}_i \mathbf{L}_i^H$, where $\mathbf{L}_i$ is a lower triangular matrix (whose inverse will be used to whiten the noise at the receive side), we obtain the following equation (using the property $det(\mathbf{I}+\mathbf{AB}) = det(\mathbf{I}+\mathbf{BA})$):

$$\mathcal{L}(\mathbf{\Lambda}, \tilde{\mathbf{\Lambda}}_1, \ldots, \tilde{\mathbf{\Lambda}}_{N_c}, (\mathbf{\Phi}_i)_{i=1...N_c}) =$$
$$\sum_{i=1}^{N_c} \left( log_2 \left[ det \left( \mathbf{I} + \mathbf{L}_i^{-1} \mathbf{H}_i \mathbf{\Phi}_i \mathbf{H}_i^H \mathbf{L}_i^{-H} \right) \right] \right.$$
$$\left. - Trace((\mathbf{\Lambda} + \tilde{\mathbf{\Lambda}}_i)\mathbf{\Phi}_i) \right) + Trace\left(\mathbf{\Lambda} diag(P_j^{tot})\right) \quad (17)$$
$$+ \sum_{i=1}^{N_c} Trace\left(\tilde{\mathbf{\Lambda}}_i diag(\phi_i^{mask,j})\right)$$

The Singular Value Decomposition (SVD) of the whitened channel $\mathbf{L}_i^{-1} \mathbf{H}_i (\mathbf{\Lambda} + \tilde{\mathbf{\Lambda}}_i)^{-1/2} = \mathbf{U}_i \mathbf{D}_i \mathbf{V}_i^H$ transforms the initial channel into a product between two unitary matrices $\mathbf{U}_i, \mathbf{V}_i^H$ and a diagonal matrix of singular values $\mathbf{D}_i$.

$$\mathcal{L}(\mathbf{\Lambda}, \tilde{\mathbf{\Lambda}}_1, \ldots, \tilde{\mathbf{\Lambda}}_{N_c}, (\mathbf{\Phi}_i)_{i=1...N_c}) =$$
$$\sum_{i=1}^{N_c} \left( log_2[det(\mathbf{I} + \mathbf{U}_i \mathbf{D}_i \mathbf{V}_i^H (\mathbf{\Lambda} + \tilde{\mathbf{\Lambda}}_i)^{1/2} \mathbf{\Phi}_i \right.$$
$$(\mathbf{\Lambda} + \tilde{\mathbf{\Lambda}}_i)^{1/2} \mathbf{V}_i \mathbf{D}_i \mathbf{U}_i^H)] - Trace(\mathbf{V}_i^H (\mathbf{\Lambda} + \tilde{\mathbf{\Lambda}}_i)^{1/2} \mathbf{\Phi}_i \quad (18)$$
$$\left. (\mathbf{\Lambda} + \tilde{\mathbf{\Lambda}}_i)^{1/2} \mathbf{V}_i) \right) + Trace\left(\mathbf{\Lambda} diag(P_j^{tot})\right)$$
$$+ \sum_{i=1}^{N_c} Trace\left(\tilde{\mathbf{\Lambda}}_i diag(\phi_i^{mask,j})\right)$$

By setting $\tilde{\mathbf{\Phi}}_i = \mathbf{V}_i^H (\mathbf{\Lambda}+\tilde{\mathbf{\Lambda}}_i)^{1/2} \mathbf{\Phi}_i (\mathbf{\Lambda}+\tilde{\mathbf{\Lambda}}_i)^{1/2} \mathbf{V}_i$ we can rewrite (18) as:

$$\mathcal{L}(\mathbf{\Lambda}, \tilde{\mathbf{\Lambda}}_1, \ldots, \tilde{\mathbf{\Lambda}}_{N_c}, (\tilde{\mathbf{\Phi}}_i)_{i=1...N_c}) = \sum_{i=1}^{N_c} \left( log_2 \left[ det \left( \mathbf{I} + \mathbf{D}_i^2 \tilde{\mathbf{\Phi}}_i \right) \right] \right.$$
$$\left. - Trace(\tilde{\mathbf{\Phi}}_i) \right) + Trace\left(\mathbf{\Lambda} diag(P_j^{tot})\right)$$
$$+ \sum_{i=1}^{N_c} Trace\left(\tilde{\mathbf{\Lambda}}_i diag(\phi_i^{mask,j})\right) \quad (19)$$

In order to find the maximum, we compute the derivative of the Lagrangian:

$$\frac{d\mathcal{L}(\mathbf{\Lambda}, \tilde{\mathbf{\Lambda}}_1, \ldots, \tilde{\mathbf{\Lambda}}_{N_c}, (\mathbf{\Phi}_i)_{i=1...N_c})}{d\tilde{\mathbf{\Phi}}_i} = \frac{1}{ln(2)} diag\left[\left(\mathbf{D}_i^{-2} + \tilde{\mathbf{\Phi}}_i\right)^{-1}\right] - \mathbf{I} = \mathbf{0} \quad (20)$$

The optimal $\mathbf{\Phi}_i$ is given by:

$$\mathbf{\Phi}_i = (\mathbf{\Lambda} + \tilde{\mathbf{\Lambda}}_i)^{-1/2} \mathbf{V}_i \left[\frac{\mathbf{I}}{ln(2)} - \mathbf{D}_i^{-2}\right]^+ \mathbf{V}_i^H (\mathbf{\Lambda} + \tilde{\mathbf{\Lambda}}_i)^{-1/2} \quad (21)$$

where the $[.]^+$ operation is inserted in order to obtain positive semi-definite $\mathbf{\Phi}_i$'s in formula (13). One can note that the precoder formulas are a function of the Lagrange multipliers $\mathbf{\Lambda}, \tilde{\mathbf{\Lambda}}_1, \ldots, \tilde{\mathbf{\Lambda}}_{N_c}$'s. This is the new generalized SVD-based closed form solution for MIMO systems under per-modem total power constraints and spectral mask constraints[2].

### B. Optimal Tx/Rx structure

The Tx/Rx structure is obtained as follows. The first step is to find the optimal Lagrange multipliers defined for the per-modem total power constraint and the spectral mask constraints according to the dual objective function $F(\Lambda, \tilde{\Lambda}_1, \ldots, \tilde{\Lambda}_{N_c})$. As the function is continuous differentiable, the search algorithm can use a gradient-descent like method to find the optimal Lagrange multipliers and is guaranteed to converge. The algorithm tries to converge under the per-modem total power constraints over the tones and inside this optimization tries to converge on a per-tone basis to also satisfy the spectral mask constraints. The complete algorithm of power allocation under per-modem total power constraints and spectral mask constraints is given in the Annex B. After calculating the optimal Lagrange multipliers, we can calculate for each tone the SVD of the whitened channel scaled by the Lagrange multipliers $\mathbf{L}_i^{-1}\mathbf{H}_i(\Lambda_{opt} + \tilde{\Lambda}_{i,opt})^{-1/2} = \mathbf{U}_i \mathbf{D}_i \mathbf{V}_i^H$ and multiply the transmitted symbols by $(\Lambda_{opt} + \tilde{\Lambda}_{i,opt})^{-1/2}\mathbf{V}_i$ and the received symbols by $\mathbf{U}_i^H$ leading to:

$$\mathbf{U}_i^H \mathbf{L}_i^{-1} \mathbf{y}_i = \mathbf{U}_i^H \mathbf{L}_i^{-1} \mathbf{H}_i (\Lambda_{opt} + \tilde{\Lambda}_{i,opt})^{-1/2} \mathbf{V}_i \tilde{\mathbf{x}}_i + \mathbf{U}_i^H \mathbf{L}_i^{-1} \mathbf{n}_i \quad (22)$$

This leads to parallel SISO systems as defined by:

$$\mathbf{U}_i^H \mathbf{L}_i^{-1} \mathbf{y}_i = \mathbf{D}_i \tilde{\mathbf{x}}_i + \mathbf{U}_i^H \mathbf{L}_i^{-1} \mathbf{n}_i \quad (23)$$

with the optimal power allocation under per-modem total power constraints and spectral mask constraints driven by (21).

### C. Per-modem total power constraints

In this paragraph we recall the main steps for finding the optimal power allocation and optimal Tx/Rx structures with self crosstalk and external noise under per-modem total power constraints. We give the SVD-based algorithm under per-modem total power constraints for the readers who are interested in the derivations without referring to the more complicated SVD-based algorithm under per-modem total power constraints and spectral mask constraints[3]. The primal problem of finding optimal PSD's for a MIMO binder under per-modem total power constraints $P_j^{tot}$ is:

$$\begin{array}{c} \max_{(\Phi_i)_{i=1\ldots N_c}} C(\Phi_i)_{i=1\ldots N_c} \\ \text{subject to } \sum_{i=1}^{N_c} [\Phi_i]_{jj} \leq P_j^{tot} \; \forall j \\ \Phi_i \succeq 0, i = 1\ldots N_c \end{array} \quad (24)$$

The dual objective function is:

$$F(\Lambda) = \max_{(\Phi_i)_{i=1\ldots N_c}} \mathcal{L}(\Lambda, (\Phi_i)_{i=1\ldots N_c}) \quad (25)$$

with $\Lambda$ a diagonal matrix of Lagrange multipliers $diag(\lambda_1, \ldots, \lambda_N)$ and

$$\Phi_i = (\Lambda + \tilde{\Lambda}_i)^{-1/2} \mathbf{V}_i \left[ \frac{\mathbf{I}}{ln(2)} - \Gamma \mathbf{D}_i^{-2} \right]^+ \mathbf{V}_i^H (\Lambda + \tilde{\Lambda}_i)^{-1/2}$$

---

[2]For practical implementations, we again introduce the SNR gap $\Gamma$ referred as the code gap in [12] which is the SNR multiplier required to achieve the target probability of error at the desired data rate. Considering the same $\Gamma$ for the different virtual channels, the optimal $\Phi_i$ is given by:

[3]The optimal power allocation under per-modem total power constraints can be found directly from (21) by setting $\tilde{\Lambda}_1, \ldots, \tilde{\Lambda}_{N_c} = \mathbf{0}$. Moreover, the optimal power allocation under a total power constraint can be found directly from (21) by setting $\tilde{\Lambda}_1, \ldots, \tilde{\Lambda}_{N_c} = \mathbf{0}$ and $\Lambda = \lambda \mathbf{I}$

$$\mathcal{L}(\Lambda, (\Phi_i)_{i=1\ldots N_c}) = \sum_{i=1}^{N_c} \left( log_2 \left[ det \left( \mathbf{I} + \mathbf{H}_i \Phi_i \mathbf{H}_i^H \mathbf{R}_i^{-1} \right) \right] \right. \\ \left. - Trace(\Lambda \Phi_i) \right) + Trace\left( \Lambda diag(P_j^{tot}) \right) \quad (26)$$

The dual optimization problem is:

$$\begin{array}{c} \underset{\Lambda}{\text{minimize}} \quad F(\Lambda) \\ \text{subject to} \quad \lambda_j \geq 0 \quad \forall j \end{array} \quad (27)$$

The search for the optimal $\Lambda$ involves evaluations of the dual objective function, i.e. maximizations of the Lagrangian, which is decoupled over the tones for a given set $\lambda_j$'s. By exploiting the Cholesky decomposition $\mathbf{R}_i = \mathbf{L}_i \mathbf{L}_i^H$, by defining the ($\Lambda$-dependent) SVD $\mathbf{L}_i^{-1}\mathbf{H}_i \Lambda^{-1/2} = \mathbf{U}_i \mathbf{D}_i \mathbf{V}_i^H$ and by setting $\tilde{\Phi}_i = \mathbf{V}_i^H \Lambda^{1/2} \Phi_i \Lambda^{1/2} \mathbf{V}_i$, we can reformulate the optimization problem as:

$$\mathcal{L}(\Lambda, (\tilde{\Phi}_i)_{i=1\ldots N_c}) = \sum_{i=1}^{N_c} \left( log_2 \left[ det \left( \mathbf{I} + \mathbf{D}_i^2 \tilde{\Phi}_i \right) \right] \right. \\ \left. - Trace(\tilde{\Phi}_i) \right) + Trace\left( \Lambda diag(P_j^{tot}) \right) \quad (28)$$

We compute the derivative of the function in order to find the maximum, therefore the optimal power allocation is given by:

$$\Phi_i = \Lambda^{-1/2} \mathbf{V}_i \left[ \frac{\mathbf{I}}{ln(2)} - \mathbf{D}_i^{-2} \right]^+ \mathbf{V}_i^H \Lambda^{-1/2} \quad (29)$$

The complete algorithm of power allocation under per-modem total power constraints is given in the Annex C. After calculating the optimal Lagrange multipliers, we can calculate for each tone the SVD of the whitened channel scaled by the Lagrange multipliers $\mathbf{L}_i^{-1}\mathbf{H}_i \Lambda_{opt}^{-1/2} = \mathbf{U}_i \mathbf{D}_i \mathbf{V}_i^H$, where $\Lambda_{opt}$ is the optimal setting for the Lagrange multipliers, we multiply the transmitted symbols by $\Lambda_{opt}^{-1/2}\mathbf{V}_i$ and the received symbols by $\mathbf{U}_i^H$ leading to:

$$\mathbf{U}_i^H \mathbf{L}_i^{-1} \mathbf{y}_i = \mathbf{U}_i^H \mathbf{L}_i^{-1} \mathbf{H}_i \Lambda_{opt}^{-1/2} \mathbf{V}_i \tilde{\mathbf{x}}_i + \mathbf{U}_i^H \mathbf{L}_i^{-1} \mathbf{n}_i \quad (30)$$

This leads to parallel SISO systems as defined by:

$$\mathbf{U}_i^H \mathbf{L}_i^{-1} \mathbf{y}_i = \mathbf{D}_i \tilde{\mathbf{x}}_i + \mathbf{U}_i^H \mathbf{L}_i^{-1} \mathbf{n}_i \quad (31)$$

with the optimal power allocation under per-modem total power constraints driven by (29).

### D. Remark

The power allocation problem under a total power constraint and spectral mask constraints can be solved similarly to the problem of power allocation under per-modem total power constraints and spectral mask constraints. In this case $\Lambda = \lambda \mathbf{I}$. The same derivation as in the previous section can be given. The optimal power allocation under total power constraint is then given by:

$$\Phi_i = (\lambda \mathbf{I} + \tilde{\Lambda}_i)^{-1/2} \mathbf{V}_i \left[ \frac{\mathbf{I}}{ln(2)} - \mathbf{D}_i^{-2} \right]^+ \mathbf{V}_i^H (\lambda \mathbf{I} + \tilde{\Lambda}_i)^{-1/2} \quad (32)$$

The PSD's formula and the algorithm description can be easily modified accordingly.

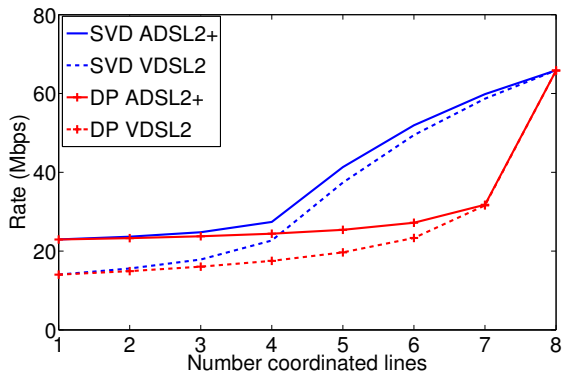

Fig. 1. Rates obtained in a downlink scenario of the SVD-based algorithm with per-modem total power constraint and the Diagonalizing Precoder for ADSL2+ and VDSL2 disturbing lines

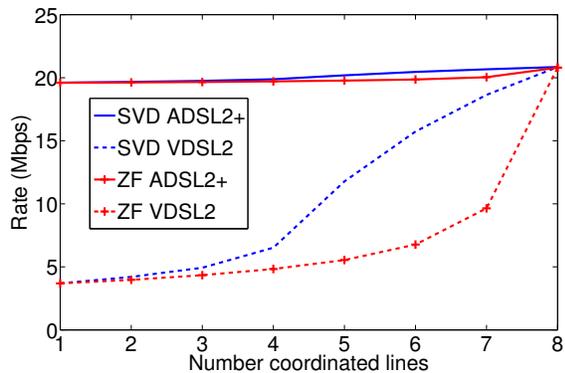

Fig. 2. Rates obtained in an uplink scenario of the SVD-based algorithm with per-modem total power constraint and the ZF receiver for ADSL2+ and VDSL2 disturbing lines

## IV. RESULTS

Far End Crosstalk (FEXT) and Near End Crosstalk (NEXT) models are well known in the literature for xDSL [7]. The NEXT and FEXT models for ADSL/HDSL may be no longer applicable to VDSL2 due to much larger bandwidth. In this paper, the simulations results are obtained on measured channels from a France Telecom binder with 8 lines of 800 meters and external noise coming from 400 meters lines. We look at the performance of VDSL2 lines with two different sources of external noise. The two sources of external noise consist of ADSL2+ or VDSL2 disturbing lines having their PSD's at spectral masks [13], [14]. This gives an indication of the statistical variations of the achievable bit-rates of the presented algorithms because different combinations of pairs may be bundled. Recently, crosstalk channels for VDSL2 have been characterized as parametric models [15]. Therefore, the optimal Tx/Rx structures and power allocation algorithms proposed in this article can also be used in such generic models[4].

We use spectral masks for VDSL2 Fiber To The exchange (FTTex) as described in [14], with SNR gap $\Gamma$=10.8 dB (Shannon gap=9.8 dB, margin=6 dB and coding gain=5 dB) to achieve the target BER, an AWGN of -140 dBm/Hz and maximum transmit power $P_j^{tot}$=14.5 dBm per line. The power spectrum of the disturbing system (ADSL2+ or VDSL2) is set to its spectral mask [13], [14]. The frequency range is from 0 to 12 MHz with 4.3125 kHz spacing between subcarriers and 4 kHz symbol rate. The FDD band plan of VDSL2 corresponds to 2 frequency bands in the downlink scenario which are 138kHz-3.75MHz and 5.2MHz-8.5MHz. In the uplink scenario, this corresponds to 3 frequency bands 25kHz-138kHz, 3.75MHz-5.2MHz and 8.5MHz-12MHz. The processing of the bundled systems is the same for both cases.

**Fig.1** shows the comparison between the two-sided coordination vector channels SVD-based algorithm and the one-sided coordination BC Diagonalizing Precoder (DP) algorithm with per-modem total power constraint in a downlink scenario [4]. The BC DP corresponds to a scaled version of the ZF precoder by the diagonal elements of the channel matrix. The length of the bonded lines are 800 meters while the coupling between the bonded lines and the disturbing lines occurs the last 400 meters to the Customer Premise Equipment (CPE). ADSL2+ disturbing lines and VDSL2 disturbing lines whose PSD's are set to their respective spectral masks are simulated. A MIMO binder of 8 lines is used, with the number of coordinated pairs going from 1 to 8 and the number of disturbing lines from 7 to 0 respectively. In each case, all $\binom{8}{N}$ combinations are used to provide an average bit-rate. When the number of coordinated pairs equals to 8, there is no external noise and thus this provides maximum performance. The SVD-based transmission with per-modem total power constraint performs better than the DP with per-modem constraint owing to the exploitation of the noise correlation by whitening. Moreover, the higher the number of coordinated pairs, the higher the improvement in terms of bit-rate. In fact, the whitening process provides more cancellation performance of the disturbing lines when the number of coordinated pairs is higher than the number of disturbing lines. Although the spectral mask of ADSL2+ is much higher than VDSL2 for frequencies up to 2.2 MHz, the VDSL2 disturbing lines have much more impact on the considered bonded lines because the crosstalk increases as the frequency increases.

We have simulated the one-sided coordination BC Minimum Mean Square Error-Dirty Paper Coding (MMSE-DPC) algorithm in a downlink scenario with optimal power allocation found by exhaustive search without external noise cancellation since noise whitening is not possible at the transmit side [2]. For the 2 user case the $\binom{8}{2}$ channels have been processed and they lead to an average bit-rate of 14.92 Mbps in the case of VDSL2 disturbing lines compared to 15.60 Mbps for the SVD-based solution with per-modem total power constraints. Therefore the external noise decreases the performance of the one-sided BC MMSE-DPC algorithm in a downlink scenario. The SVD-based algorithm provides an upper bound on the achievable capacity, thus it exploits the equivalent whitened channel matrix where optimal power allocation is found by a closed form formula contrary to one-sided coordination BC MMSE-DPC where optimal power allocation is found by exhaustive search.

**Fig.2** shows the comparison between the two-sided coordination vector channels SVD-based algorithm and the one-sided coordination MAC ZF receiver with per-modem total power constraint in an uplink scenario [3]. Similar conclusions can be told for this comparison, however even if the whitening step is possible at the receive side, the ZF equalizer can't take advantage of the equivalent channel as the SVD scheme does. Indeed, the ZF structure consists of an inversion of the channel matrix, and the performance is not changed by inverting the whitened channel. Moreover, there is a very small difference between ZF and SVD schemes for ADSL2+ crosstalkers because they don't harm the VDSL2 lines. Indeed, the bandwidth involved in the uplink scenario is the 25kHz-138kHz bandwidth, therefore SISO schemes with 7 ADSL2+ crosstalkers and 8x8 MIMO schemes with no ADSL2+ crosstalkers have almost the same performance. Contrary to ADSL2+ crosstalkers, VDSL2 crosstalkers decrease the performance of the VDSL2 lines. For the 2 user case, the one-sided coordination MAC MMSE-Decision Feedback Equalizer (MMSE-DFE) algorithm leads to an average bit-rate of 4.20 Mbps in the case of VDSL2 disturbing lines, which is similar to the average bit-rate of the SVD-based algorithm. In this case, the whitening is possible at the receive side and the MMSE-DFE receiver can exploit the equivalent channel.

---

[4]Note that from the implementation point of view, once the optimal covariance matrices $(\mathbf{\Phi}_i)_{i=1...N_c}$ are determined, the transmitted data symbols $\mathbf{x}_i$ can be constructed as follows:
1) The $N \times 1$ vector of the M-QAM data symbols $\mathbf{s}_i$ is precoded using the $N \times N$ $\mathbf{C}_i$ matrix, i.e. $\mathbf{x}_i = \mathbf{C}_i \mathbf{s}_i$, such that $E[\mathbf{x}_i \mathbf{x}_i^H] = \mathbf{\Phi}_i = \mathbf{C}_i \mathbf{C}_i^H$ from Cholesky decomposition (as $\mathbf{\Phi}_i$ is a positive semi-definite matrix).
2) Then $\mathbf{x}_i$ will be sent on the $N$ lines (the $j^{\text{th}}$ element of $\mathbf{x}_i$ will be sent on the $j^{\text{th}}$ line).

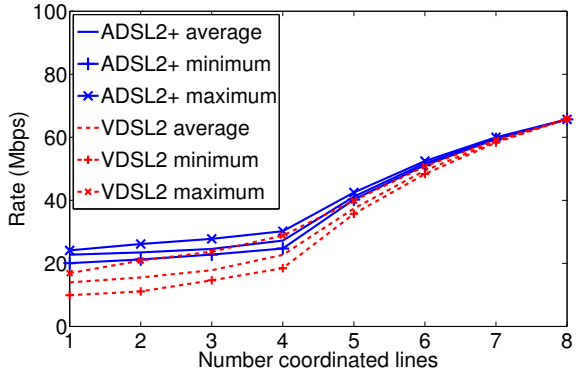

Fig. 3. Rates obtained in a downlink scenario for the different permutations of SVD algorithm under per-modem total power constraints and spectral mask constraints for ADSL2+ and VDSL2 disturbing lines

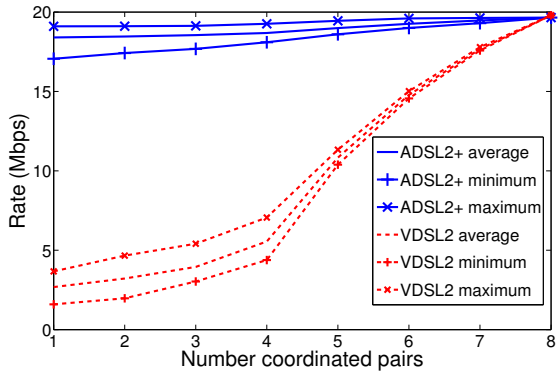

Fig. 4. Rates obtained in an uplink scenario for the different permutations of SVD algorithm under per-modem total power constraints and spectral mask constraints for ADSL2+ and VDSL2 disturbing lines

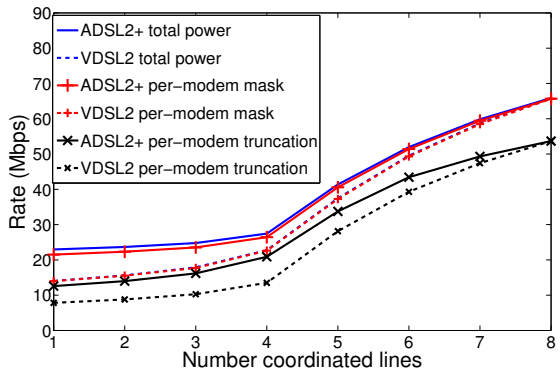

Fig. 5. Rates obtained in a downlink scenario of the different SVD-based algorithms under total power constraint, per-modem total power constraints and spectral mask constraints and per-modem total power constraints using (a posteriori) truncation by the spectral mask for ADSL2+ and VDSL2 disturbing lines

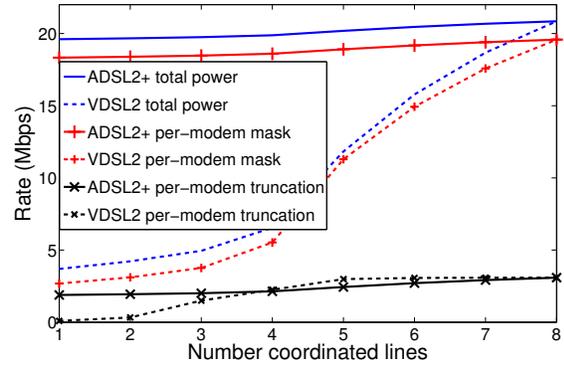

Fig. 6. Rates obtained in an uplink scenario of the different SVD-based algorithms under total power constraint, per-modem total power constraints and spectral mask constraints and per-modem total power constraints using (a posteriori) truncation by the spectral mask for ADSL2+ and VDSL2 disturbing lines

**Fig.3** and **Fig.4** show the bit-rate performance of the two-sided coordination vector channels SVD algorithm under per-modem total power constraints and spectral mask constraints for the downlink and the uplink scenario with ADSL2+ and VDSL2 disturbing lines. The $\binom{8}{N}$ combinations are used to provide an average bit-rate, a minimum bit-rate and a maximum bit-rate, thus variations in bit-rates using all different permutations. One can observe that the range of bit-rates diminishes as the number of coordinated lines increases. The comments given in **Fig.1** and **Fig.2** also apply to these figures.

**Fig.5** and **Fig.6** give the bit-rate performance of the two-sided coordination vector channels SVD-based algorithms under a total power constraint, per-modem total power constraints using (a posteriori) truncation by the spectral mask[5], per-modem total power constraints and spectral mask constraints for the downlink and the uplink scenario with ADSL2+ and VDSL2 disturbing lines. A total power constraint gives an extra degree of freedom compared to per-modem total power constraints and thus achieves better performance. Similar comments from the previous figures could be told. We observe that there are very small differences between bit-rates under a total power constraint and bit-rates under per-modem total power constraints. For the SVD algorithm with per-modem total power constraints using (a posteriori) truncation by the spectral mask, the spectral masks are directly applied to the optimal PSD's under per-modem total power constraints. There is a noticeable difference between the performance of the SVD algorithm using truncation by spectral mask and the SVD algorithm under spectral mask constraints especially in the uplink scenario. In fact, this optimization process allows a better distribution of the power over the tones by setting the optimal PSD's larger than the spectral mask at the spectral mask and thus saving power for other tones.

---

[5]The truncation by the spectral mask cuts the optimal PSD's found by the SVD-based algorithm under per-modem total power constraints and do not distribute to further tones the power loss due to the truncation when the PSD's are higher than the spectral mask

We can notice that the gain of the proposed techniques originates from the MIMO diversity gain and does not originates from waterfilling on the considered loops. In fact, the eigenvalues used in the SVD-based algorithms do not have a significant impact on the optimal power allocation. The inverse of eigenvalues is in the order of $10^{-8}$ and the water level is around $10^{-6}$ for the considered loops and the total power constraint in (11) (the same behaviour can be observed under per-modem total power constraints and mask constraints in (21)). As the heart of the waterfilling solution reduces the water level by the inverse of the eigenvalues, we observe only small variations around the water level and the waterfilling operation leads to flat PSD's. Destroying this diagonal dominance by increasing artificially the crosstalk before the withening operation, results in a capacity gain originating both from MIMO diversity and the waterfilling gains. The resulted optimal power allocation (not included in the manuscript for the space limitations) is no longer flat for the two lines. This suggests that these SVD-based algorithms could have a significant impact in the wireless context where the channel matrix is not diagonally dominant.

In general, in VDSL2 scenario's with a binder with equal length cables, we can expect that per-modem total power constraints and spectral mask constraints do not degrade performance, while these constraints do lead to more practical, implementable and standards-compliant solutions. Moreover, power allocation algorithms under spectral mask constraints can provide better performance compared to a simpler power allocation procedure where the spectral mask constraints are first removed from the optimization problem, and then imposed onto the computed PSD's. This is especially so in uplink scenario's or even more in a scenario where optimal PSD's could be much larger than spectral mask PSD's.

## V. CONCLUSION

In this paper, we have investigated the problem of finding the optimal power allocation in MIMO xDSL systems under self crosstalk and external noise and with two-sided coordination (full vectoring). Optimal Tx/Rx structures and power allocation algorithms have been devised under practical limitations from xDSL systems, namely per-modem total power constraints and spectral mask constraints, leading to a generalized SVD-based transmission. Simulation results were given for bonded VDSL2 systems with external noise coming from ADSL2+ or VDSL2 disturbing lines, along with a comparison with algorithms with one-sided signal coordination, either only at the transmit side or the receive side. The two-sided coordination SVD-based algorithms then provide a performance upper bound for the existing one-sided coordination MAC ZF, BC DP, MAC MMSE-DFE or BC MMSE-DPC algorithms under self crosstalk and external noise. The simulation results also showed that adding per-modem total power constraints and spectral mask constraints did not significantly reduce the bit-rate compared to the case where only a total power constraint is imposed owing to the SVD-based transmission. The optimal power allocation algorithms under spectral mask constraints also provides better performance compared to a simpler power allocation procedure where the spectral mask constraints are first removed from the optimization problem, and then imposed onto the computed PSD's. An extension of this work could be aimed at canceling external noise using the noise correlation between differential-mode and common-mode signals, or between different tones when considering asynchronous external noise.

## ANNEX

This Annex provide the algorithms for finding the optimal Lagrange multipliers for the two-sided coordination vector channels SVD-based algorithms under a total power constraint, per-modem total power constraints and spectral mask constraints and per-modem total power constraints respectively.

### A. Total power constraint

The following algorithm provides the optimal power allocation for the two-sided coordination vector channels under self crosstalk and external noise using an SVD-based algorithm. The later tries to find in an iterative way the optimal Lagrange multiplier to meet the total power constraint.

---
**Algorithm 1** Total power constraint
---
init $\lambda = 1$
init $step = 2$
init $b = 0$
init $\boldsymbol{\Phi}_i = \mathbf{V}_i \left[ \frac{\mathbf{I}}{ln(2)\lambda} - \mathbf{D}_i^{-2} \right]^+ \mathbf{V}_i^H \ \forall i$
while $| \sum_{i=1}^{N_c} Trace(\boldsymbol{\Phi}_i) - P^{tot} | > tolerance$
　　if $\sum_{i=1}^{N_c} Trace(\boldsymbol{\Phi}_i) - P^{tot} < 0$
　　　　$b = b + 1$
　　　　$\lambda = \lambda / step$
　　　　$step = step - 1/2^b$
　　end if
　　$\lambda = \lambda * step$
　　$\boldsymbol{\Phi}_i = \mathbf{V}_i \left[ \frac{\mathbf{I}}{ln(2)\lambda} - \mathbf{D}_i^{-2} \right]^+ \mathbf{V}_i^H \ \forall i$
end while

---

### B. Per-modem total power constraints and spectral mask constraints

The following algorithm provides the optimal power allocation for the two-sided coordination vector channels under self crosstalk and external noise using an SVD-based algorithm. The later tries to find in an iterative way the optimal Lagrange multipliers to meet per-modem total power constraints and spectral mask constraints.

**Algorithm 2** Per-modem total power constraints and spectral mask constraints

init $\lambda_j = 1\ \forall j$
init $step_j = 2\ \forall j$
init $b_j = 0\ \forall j$
init $\mathbf{\Phi}_i = (\mathbf{\Lambda} + \tilde{\mathbf{\Lambda}_i})^{-1/2}\mathbf{V}_i \left[\frac{\mathbf{I}}{ln(2)} - \mathbf{D}_i^{-2}\right]^+ \mathbf{V}_i^H (\mathbf{\Lambda} + \tilde{\mathbf{\Lambda}_i})^{-1/2}\ \forall i$
while $|\sum_{i=1}^{N_c}[\mathbf{\Phi}_i]_{jj} - P_j^{tot}| < tolerance\ \exists j$
    for i=1 to $N_c$
        init $\tilde{\lambda}_j = 1\ \forall j$
        init $\tilde{step}_j = 2\ \forall j$
        init $\tilde{b}_j = 0\ \forall j$
        while $\tilde{b}_j < iterations$
            for j=1 to N
                $\mathbf{\Phi}_i = (\mathbf{\Lambda} + \tilde{\mathbf{\Lambda}_i})^{-1/2}\mathbf{V}_i \left[\frac{\mathbf{I}}{ln(2)} - \mathbf{D}_i^{-2}\right]^+ \mathbf{V}_i^H (\mathbf{\Lambda} + \tilde{\mathbf{\Lambda}_i})^{-1/2}\ \forall i$
                if $[\mathbf{\Phi}_i]_{jj} > \phi_i^{mask}\ \forall i$
                    $[\mathbf{\Phi}_i]_{jj} = \phi_i^{mask}\ \forall i$
                end if
                if $[\mathbf{\Phi}_i]_{jj} - \phi_i^{mask} < 0\ \forall i$
                    $\tilde{b}_j = \tilde{b}_j + 1$
                    $\tilde{\lambda}_j = \tilde{\lambda}_j / \tilde{step}_j$
                    $\tilde{step}_j = \tilde{step}_j - 1/2^{\tilde{b}_j}$
                end if
                $\tilde{\lambda}_j = \tilde{\lambda}_j * \tilde{step}_j$
            end for
        end while
    end for
    for j=1 to N
        if $\sum_{i=1}^{N_c}[\mathbf{\Phi}_i]_{jj} - P_j^{tot} < 0$
            $b_j = b_j + 1$
            $\lambda_j = \lambda_j / step_j$
            $step_j = step_j - 1/2^{b_j}$
        end if
        $\lambda_j = \lambda_j * step_j$
    end for
    $\mathbf{\Phi}_i = (\mathbf{\Lambda} + \tilde{\mathbf{\Lambda}_i})^{-1/2}\mathbf{V}_i \left[\frac{\mathbf{I}}{ln(2)} - \mathbf{D}_i^{-2}\right]^+ \mathbf{V}_i^H (\mathbf{\Lambda} + \tilde{\mathbf{\Lambda}_i})^{-1/2}\ \forall i$
end while

**Algorithm 3** Per-modem total power constraints

init $\lambda_j = 1\ \forall j$
init $step_j = 2\ \forall j$
init $b_j = 0\ \forall j$
init $\mathbf{\Phi}_i = \mathbf{\Lambda}^{-1/2}\mathbf{V}_i \left[\frac{\mathbf{I}}{ln(2)} - \mathbf{D}_i^{-2}\right]^+ \mathbf{V}_i^H \mathbf{\Lambda}^{-1/2}\ \forall i$
while $|\sum_{i=1}^{N_c}[\mathbf{\Phi}_i]_{jj} - P_j^{tot}| > tolerance\ \exists j$
    for j=1 to N
        if $\sum_{i=1}^{N_c}[\mathbf{\Phi}_i]_{jj} - P_j^{tot} < 0$
            $b_j = b_j + 1$
            $\lambda_j = \lambda_j / step_j$
            $step_j = step_j - 1/2^{b_j}$
        end if
        $\lambda_j = \lambda_j * step_j$
    end for
    $\mathbf{\Phi}_i = \mathbf{\Lambda}^{-1/2}\mathbf{V}_i \left[\frac{\mathbf{I}}{ln(2)} - \mathbf{D}_i^{-2}\right]^+ \mathbf{V}_i^H \mathbf{\Lambda}^{-1/2}\ \forall i$
end while

*C. Per-modem total power constraints*

The following algorithm provides the optimal power allocation for the two-sided coordination vector channels under self crosstalk and external noise using an SVD-based algorithm. The later tries to find in an iterative way the optimal Lagrange multipliers to meet per-modem total power constraints.